\documentstyle[12pt,amssymb]{article}
\begin{document}

\title{Non-commutative space-time and the uncertainty principle}
\date{}
\author{Eric Carlen\thanks{%
School of Mathematics, Georgia Institute of Technology, Atlanta GA 30332
USA, carlen@math.gatech.edu;\ Work partially supported by NSF grant DMS
00-70589}\quad and R. Vilela Mendes\thanks{%
Grupo de F\'{i}sica Matem\'{a}tica, Universidade de Lisboa, Av. Gama Pinto
2, 1699 Lisboa, Portugal, vilela@cii.fc.ul.pt}}
\maketitle

\begin{abstract}
The full algebra of relativistic quantum mechanics (Lorentz plus Heisenberg)
is unstable. Stabilization by deformation leads to a new deformation
parameter $\varepsilon \ell ^{2}$, $\ell $ being a length and $\varepsilon $
a $\pm $ sign. The implications of the deformed algebras for the uncertainty
principle and the density of states are worked out and compared with the
results of past analysis following from gravity and string theory.
\end{abstract}


Physical theories are approximations to Nature and physical constants may
never be known with absolute precision. Therefore, a wider range of validity
is expected to hold for theories that do not change in a qualitative manner
under a small change of parameters. Such theories are called {\it stable} or 
{\it rigid}. The stable-model point of view originated in the field of
non-linear dynamics, where it led to the notion of {\it structural stability}%
\cite{Andronov} \cite{Smale}. However, as emphasized by Flato\cite{Flato}
and Faddeev\cite{Faddeev}, the same pattern seems to occur in the
fundamental theories of Nature. Indeed, the most important physical
revolutions of this century, the transition from non-relativistic to
relativistic and from classical to quantum mechanics, may be interpreted as
the replacement of two unstable theories by two stable ones. Mathematically
this corresponds to stabilizing deformations leading, in the first case,
from the Galilean to the Lorentz algebra and, in the second, from the
algebra of commutative phase-space to the Moyal-Vey algebra (or equivalently
to the Heisenberg algebra). The deformation parameters, which for non-zero
values make the algebras stable, are $\frac{1}{c}$ (the inverse of the speed
of light) and $h$ (the Planck constant). Once deformed, the algebras are all
equivalent for non-zero values of $\frac{1}{c}$ and $h$. Hence, relativistic
mechanics and quantum mechanics may be derived from the conditions for
stability of two mathematical structures, although the exact values of the
deformation parameters cannot be fixed by purely algebraic considerations.
Instead, the deformation parameters are fundamental constants to be obtained
from experiment. In this sense not only is deformation theory the theory of
stable theories, it is also the theory that identifies the fundamental
constants.

Some time ago it was noticed\cite{Vilela1} that stability of the subalgebras
does not guarantee stability of the full algebra of relativistic quantum
mechanics. The latter contains the Lorentz algebra $\{M_{\mu \nu }\}$ and
the Heisenberg algebras $\left\{ p_{\mu },x_{\nu }\right\} $ plus the
commutators that define the 4-vector nature of $p_{\mu }$ and $x_{\mu }$.
The full algebra turns out to be unstable and its stabilization by
deformation leads to a new deformation parameter $\ell ^{2}$ with dimension
(length)$^{2}$. The deformed commutators are 
\begin{equation}
\begin{array}{rcl}
\lbrack x_{\mu },x_{\nu }] & = & -i\varepsilon \ell ^{2}M_{\mu \nu } \\ 
\lbrack p_{\mu },x_{\nu }] & = & i\eta _{\mu \nu }\Im \\ 
\lbrack x_{\mu },\Im ] & = & i\varepsilon \ell ^{2}p_{\mu }
\end{array}
\label{1}
\end{equation}
where $\eta _{\mu \nu }=(1,-1,-1,-1)$, $c=\hbar =1$ , $\varepsilon =\pm 1$
and $\Im $ is the operator that replaces the trivial center of the
Heisenberg algebra. The new relativistic quantum mechanics algebra implies
that space time is a non-commutative manifold and has other physical
consequences, some of which are explored in Refs.\cite{Vilela2}, \cite
{Vilela3}. Also, like $\frac{1}{c}$ and $h$, the deformation parameter $\ell 
$ is naturally identified as a new fundamental constant to be obtained from
experiment. This constant sets the scale for the spectrum of the position
operators. In addition to the magnitude $\ell $, there is also the sign $%
\varepsilon $ of the deformation parameter that is not fixed by stability
considerations and must be determined experimentally.

Notice that because of non-commutativity of the space-time coordinates only
one coordinate may be sharply specified. For the choice $\varepsilon =-1$
the space coordinates have discrete spectrum but the time coordinate a
continuous spectrum and conversely for $\varepsilon =+1$\cite{Vilela3}. We
focus here on the one--dimensional subalgebras (for one space coordinate and
one momentum) that replace Heisenberg's algebra, which are 
\begin{equation}
\begin{array}{lll}
\left[ x,p\right] & = & i\Im \\ 
\left[ x,\Im \right] & = & i\varepsilon \ell ^{2}p \\ 
\left[ p,\Im \right] & = & 0
\end{array}
\label{2}
\end{equation}
with either $\varepsilon =-1$ or $\varepsilon =+1$.

For $\varepsilon =-1$ this is the algebra of the group of motions of the
plane, ISO(2), and for $\varepsilon =+1$ the algebra of the group of motions
of the hyperbolic plane, ISO(1,1).

For the ISO(2) case the irreducible representations $T_{r}$ \cite{Vilenkin}
can be realized as operators on $L^{2}(S^{1})$ with respect to normalized
Lebesgue measure on $S^{1}$, so that the scalar product is given by 
\begin{equation}
\left( f_{1},f_{2}\right) =\frac{1}{2\pi }\int_{0}^{2\pi }f_{1}(\theta
)f_{2}^{*}(\theta )d\theta \ .  \label{3}
\end{equation}
In this case $p$ and $\Im $ are diagonal and the operators are 
\begin{equation}
\begin{array}{lll}
x & = & i\ell \frac{\partial }{\partial \theta } \\ 
p & = & r\frac{1}{\ell }\sin \theta \\ 
\Im & = & r\cos \theta
\end{array}
\label{4}
\end{equation}
Fourier transforming, we have a representation on $\ell ^{2}(Z)$ in which $x$
is diagonal 
\begin{equation}
\begin{array}{lll}
x & = & \ell n \\ 
p & = & \frac{1}{i\ell }\Delta _{-} \\ 
\Im & = & \Delta _{+}
\end{array}
\label{4a}
\end{equation}
$\Delta _{-}$ and $\Delta _{+}$ being the operators 
\begin{equation}
\begin{array}{lll}
\Delta _{-}f(x) & = & \frac{1}{2}\left( f(x+1)-f(x-1)\right) \\ 
\Delta _{+}f(x) & = & \frac{1}{2}\left( f(x+1)+f(x-1)\right)
\end{array}
\label{4b}
\end{equation}
The representations $T_{r}$ are infinite dimensional for all $r\neq 0$, a
convenient basis being the set of exponentials $\exp \left( -in\theta
\right) $%
\begin{equation}
\left\{ e^{-in\theta };n\in Z\right\}  \label{5}
\end{equation}
or in the Fourier transformed representation, 
\begin{equation}
\left\{ \delta _{n}\ ;n\in Z\right\} \ .  \label{5bis}
\end{equation}
The states $e^{-in\theta }$ are eigenstates of the position operator $x$,
which has a discrete spectrum $(=\ell Z$ $)$. $\ell $ is the minimal length
spacing and the maximum momentum $p$ is $r\frac{1}{\ell }$.

For each localized state $e_{n}=\frac{1}{\sqrt{2\pi }}e^{-in\theta }$, $P=%
\frac{1}{\ell }p$ is a random variable with characteristic function 
\begin{equation}
C(s)=<e_{n},e^{isP}e_{n}>=J_{0}(sr)  \label{6}
\end{equation}
the corresponding probability density being 
\begin{equation}
\begin{array}{lllll}
\nu (P) & = & \frac{1}{\pi }\frac{1}{\sqrt{r^{2}-P^{2}}} &  & \left|
P\right| <r \\ 
& = & 0 &  & \left| P\right| >r
\end{array}
\label{7}
\end{equation}

For the ISO(1,1) case ($\varepsilon =+1$) the irreducible representations $%
T_{r}$ \cite{Vilenkin} are realized as operators on the space of smooth
functions on the hyperbola ($\xi _{1}=\cosh \mu $ ; $\xi _{2}=\sinh \mu $ ; $%
\xi _{1}^{2}-\xi _{2}^{2}=1$) with scalar product 
\begin{equation}
\left( f_{1},f_{2}\right) =\int_{-\infty }^{\infty }f_{1}(\mu )f_{2}^{*}(\mu
)d\mu  \label{7a}
\end{equation}
the operators being 
\begin{equation}
\begin{array}{lll}
x & = & i\ell \frac{\partial }{\partial \mu } \\ 
p & = & r\frac{1}{\ell }\sinh \mu \\ 
\Im & = & r\cosh \mu
\end{array}
\label{7b}
\end{equation}
More details on the physical consequences of this algebra will be given
below.

The algebraic structure of non-commutative space-time and in particular the
choice of the sign of $\varepsilon $ has a strong effect on the spectrum.
This is, for example, illustrated by the behavior of the energy levels in a
strongly localizing potential. In the case $\ell =0$, one knows that in a
infinite square well of width $\Delta $ the energy levels $E_{n}$ arising
for 
\begin{equation}
-\frac{1}{2m}\frac{d^{2}}{dx^{2}}\psi =E\psi  \label{A1}
\end{equation}
with boundary conditions $\psi _{n}(-\frac{\Delta }{2})=\psi _{n}(\frac{%
\Delta }{2})=0$, are 
\begin{equation}
E_{n}=\frac{n^{2}\pi ^{2}}{2m\Delta ^{2}}  \label{A2}
\end{equation}
with $n=1,2,\cdots $. In particular, the ground state energy $E_{0}$
diverges quadratically in the sharp localization limit $\Delta \rightarrow 0$%
.

In the case $\ell \neq 0$ and $\varepsilon =+1$, the equation that replaces (%
\ref{A1}) is 
\begin{equation}
\frac{1}{2m}\frac{1}{4\ell ^{2}}\left( e^{i\ell \frac{d}{dx}}-e^{-i\ell 
\frac{d}{dx}}\right) ^{2}\psi =E\psi  \label{A3}
\end{equation}
leading, with the same boundary conditions, to the energy levels 
\begin{equation}
E_{n}=\frac{1}{2m\ell ^{2}}\sinh ^{2}\left( \frac{n\pi \ell }{\Delta }\right)
\label{A4}
\end{equation}
with $n=1,2,\cdots $. Again, the ground state energy diverges in the sharp
localization limit, but much more rapidly, and it coincides with (\ref{A2})
in the $\ell \rightarrow 0$ limit.

For $\ell \neq 0$ and $\varepsilon =-1$ the situation is rather different.
First of all, we must require that $\Delta $ be an integral multiple of $%
\ell $ in this case, say $\Delta =k\ell $. The equation is 
\begin{equation}
-\frac{1}{2m}\frac{1}{4\ell ^{2}}\left( e^{\ell \frac{d}{dx}}-e^{-\ell \frac{%
d}{dx}}\right) ^{2}\psi =E\psi  \label{A4a}
\end{equation}
with energy spectrum 
\begin{equation}
E_{n}=\frac{1}{2m\ell ^{2}}\sin ^{2}\left( \frac{n\pi \ell }{\Delta }\right)
\label{A4bb}
\end{equation}
This time the energies are all finite, bounded above by $1/(2m\ell ^{2})$,
independent of $\Delta =k\ell $. However, in this case $n$ runs only from $%
n=1$ to $n=k/2$ for a total of $k$ states.

We begin by analyzing the consequences of the deformation on phase space
volume counting rules. That is, confine $n$ fermions in a box of size $%
\Delta $ so that each one must occupy a different state. Let $p_{n}$ be the
magnitude of the momentum of the $n$th particle. Now add an $(n+1)$st
particle, whose momentum is $p_{n+1}$ in magnitude. The additional phase
space volume required to accommodate the new particle is 
\begin{equation}
\Delta p\Delta x=(p_{n+1}-p_{n})\Delta  \label{A5}
\end{equation}
since in the box $\Delta x=\Delta $.

In the case $\ell =0$, we have $p_{n}=\sqrt{2mE_{n}}=(n\pi )/\Delta $, and
so 
\begin{equation}
\Delta p=(p_{n+1}-p_{n})={\frac{\pi }{\Delta }}  \label{A6}
\end{equation}
Therefore, in the case $\ell =0$ 
\begin{equation}
\Delta p\Delta x=\pi  \label{A7}
\end{equation}
independent of $n$, which is the usual phase space volume counting rule.

For $\ell \ne 0$ and $\varepsilon =+1$, one easily computes from $p_{n}=%
\sqrt{2mE_{n}}$ that 
\begin{equation}
\Delta p=(p_{n+1}-p_{n})={\frac{2}{\ell }}\sinh \left( \frac{\pi \ell }{%
2\Delta }\right) \cosh \left( \frac{\pi \ell }{\Delta }(n+\frac{1}{2})\right)
\label{A8}
\end{equation}
Therefore, in this case, 
\begin{equation}
\Delta p\Delta x={\frac{2\Delta }{\ell }}\sinh \left( \frac{\pi \ell }{%
2\Delta }\right) \cosh \left( \frac{\pi \ell }{\Delta }(n+\frac{1}{2})\right)
\label{A9}
\end{equation}
This reduces to the previous result in the limit $\ell \rightarrow 0$, but
for $\ell \ne 0$, the required increase in phase space volume, to add
another particle, increases rapidly with $n$. When $n$ is large, this effect
can be significant even for very small values of $\ell $.

Finally, for $\ell \ne 0$ and $\varepsilon =+1$, we again fix $\Delta =k\ell 
$, and in the same way we compute that at the $n$th energy level, 
\begin{equation}
\Delta p\Delta x=2k\sin \left( \frac{\pi }{2k}\right) \cos \left( \frac{\pi 
}{k}(n+\frac{1}{2})\right)  \label{A10}
\end{equation}
Again, the phase space volume increment depends on $n$, but in such a way
that adding a particle at higher energy requires less and less phase space
volume.

Phase space volume counting plays an important role in statistical
mechanics, and it is conceivable that statistical mechanical considerations
could lead to bounds on the possible values of $\ell $ and the sign of $%
\varepsilon $.

We now turn to the implications of the deformed non-commutative space-time
algebra to the uncertainty principle and compare it with previous analysis
and conjectures concerning modifications of this principle following from
gravity and string theory.

Phase space volume counting is directly connected with the uncertainty
principle, and recently a number of authors have considered the introduction
of a fundamental length through a modified uncertainty principle. Indeed,
hints for the existence of a fundamental length had already appeared in
string theory\cite{Gross} \cite{Amati} \cite{Konishi} \cite{Yoneya}, as well
as through considerations of the effect of gravitation on the measurement
process \cite{Garay} \cite{Adler}. This has led to the proposal of a
generalized uncertainty principle 
\begin{equation}
\Delta x\geq \frac{\hbar }{2\Delta p}+\frac{C}{4}\frac{\Delta p}{\hbar }
\label{8}
\end{equation}
$C$ being a quantity proportional to the string tension or to the square of
Planck's length. If $\Delta p$ is finite, the inequality (\ref{8}) implies 
\begin{equation}
\Delta x\geq \sqrt{\frac{C}{2}}  \label{8a}
\end{equation}
that is, there would be a non-zero minimal length that can be probed with
finite energy states. On the other hand the statistical mechanics
consequences of (\ref{8}) have also been explored\cite{Lubo}.

We now wish to relate the conjectured generalized uncertainty principle (\ref
{8}) with the results following from non-commutative space-time algebraic
structure obtained by deformation theory. The algebraic structure of the
relation (\ref{8}) has been studied by a number of authors\cite{Maggiore} 
\cite{Kempf1} \cite{Kempf2}. Here we will refer in particular to the results
of Kempf, Mangano and Mann\cite{Kempf2}. The relation (\ref{8}) is shown to
follow from a commutation relation 
\begin{equation}
\left[ x,p\right] =i\left( 1+\frac{C}{2}p^{2}\right)  \label{9}
\end{equation}
In fact, from the Schwartz inequality one has ($\hbar =1$) 
\begin{equation}
\Delta x\Delta p\geq \frac{1}{2}\left| i\left( x\psi ,p\psi \right) -i\left(
p\psi ,x\psi \right) \right|  \label{10a}
\end{equation}
which, if the domain $D\left( \left[ x,p\right] \right) $ of $\left[
x,p\right] $ coincides with $D\left( x\right) \cap D\left( p\right) $ is
equivalent to 
\begin{equation}
\Delta x\Delta p\geq \frac{1}{2}\left| \left\langle \left[ x,p\right]
\right\rangle \right|  \label{10}
\end{equation}
Using (\ref{9}) the inequality (\ref{8}) follows.

For purposes of comparison with (\ref{2}), we note from (\ref{4}) and (\ref
{7b}) that with $r=1$, 
\begin{equation}
\lbrack x,p]=i\left( 1+\varepsilon (\ell p)^{2}\right) ^{1/2}\ .  \label{11a}
\end{equation}
Expanding in $\ell p$ to leading order, we obtain 
\begin{equation}
\lbrack x,p]\approx i\left( 1+{\frac{\varepsilon }{2}}(\ell p)^{2}\right)
\label{11b}
\end{equation}
which, in the case $\varepsilon =+1$, agrees with (\ref{9}) if we identify $%
C=\ell ^{2}$. Therefore the deformation parameter $\ell ^{2}$ is seen to
play the same role as the squared Planck length or the string tension that
appear in the generalized uncertainty relation (\ref{8}). However there are
some fundamental differences. The first is that the commutation relations (%
\ref{9}) do not correspond to a Lie algebra deformation. The spectral
structure is also different. To understand this, consider an explicit
symmetric operator realization of (\ref{9}) by operators in ${\Bbb {R}}$,
namely 
\begin{equation}
\begin{array}{lll}
p & = & p \\ 
x & = & i\left( 1+\frac{C}{2}p^{2}\right) \frac{d}{dp}+i\frac{C}{2}p
\end{array}
\label{13}
\end{equation}
This $x$ operator has normalizable eigenvectors 
\begin{equation}
\psi \left( p\right) =\left( 1+\frac{C}{2}p^{2}\right) ^{-\frac{1}{2}}\exp
\left( \sqrt{\frac{2}{C}}a\tan ^{-1}\left( \sqrt{\frac{C}{2}}p\right) \right)
\label{14}
\end{equation}
with $x\psi =a\psi $. However these states have infinite energy. The same
happens of course for the generalized position eigenstates in the usual
Heisenberg algebra. The important difference is that, contrary to the
Heisenberg case, here these $\psi $ states cannot be approximated,
arbitrarily close, by finite energy states\cite{Kempf2}. This is the reason
for the upper bound (\ref{8a}) on $\Delta x$.

The situation in the deformed algebra (\ref{2}), with $\varepsilon =+1$, is
different. From (\ref{7b}) it follows that in any non-trivial representation
(of the subalgebra) one has the relation 
\begin{equation}
\Im =\left( 1+\ell ^{2}p^{2}\right) ^{\frac{1}{2}}  \label{15}
\end{equation}
From this and the inequality (\ref{10}) one obtains the following
uncertainty principle 
\begin{equation}
\Delta x\Delta p\geq \frac{1}{2}\left| \left\langle \left( 1+\ell
^{2}p^{2}\right) ^{\frac{1}{2}}\right\rangle \right|   \label{16}
\end{equation}
In leading $\ell ^{2}$ order it looks like (\ref{8}), however the physical
content is somewhat different. In particular, if the uncertainty principle (%
\ref{16}) is used to compute partition functions as in \cite{Lubo}, the
phase-space measure will be $dxdp/\sqrt{1+\ell ^{2}p^{2}}$ rather than $%
dxdp/\left( 1+\beta p^{2}\right) $.

Eq.(\ref{16}) surely implies that the overall position-momentum uncertainty
grows when one probes a system with large momentum particles, as in (\ref{8}%
). However there is no lower bound on $\Delta x$ if one is ready to accept a
sufficiently large, but finite, uncertainty in $\Delta p$. This is already
seen in Eq.(\ref{A9}). As another example consider a normalized Gaussian 
\begin{equation}
\psi \left( \mu \right) =\left( 2\pi \alpha \right) ^{-\frac{1}{4}}\exp
\left( -\frac{\mu ^{2}}{4\alpha }\right)   \label{17}
\end{equation}
Using the representation (\ref{7b}) one computes the expectation values $%
\left( \psi ,x\psi \right) =\left( \psi ,x\psi \right) =0$ and 
\begin{equation}
\begin{array}{lll}
\left( \psi ,x^{2}\psi \right)  & = & \frac{\ell ^{2}}{4\alpha } \\ 
\left( \psi ,p^{2}\psi \right)  & = & \frac{1}{4\ell ^{2}}\left( 2e^{2\alpha
}-1\right) 
\end{array}
\label{18}
\end{equation}
Therefore one sees that, by increasing $\alpha $, $\Delta x$ may be made
arbitrarily small. However $\Delta p$ grows much faster than for the
Heisenberg algebra, namely 
\begin{equation}
\Delta x\Delta p=\frac{1}{4}\left( \frac{2e^{\alpha }-1}{\alpha }\right) 
\label{19}
\end{equation}
to be compared with $\Delta x\Delta p=\frac{1}{2}$ for the Heisenberg
algebra.

In conclusion: From the deformed algebra in the $\varepsilon =+1$ case one
obtains a modified uncertainty relation (\ref{16}) which contains the
expected higher uncertainty associated to large momentum probes, but no
nonzero lower bound on $\Delta x$.

For the $\varepsilon =-1$ case, as seen above, each space coordinate has a
discrete spectrum, in units of $\ell $. Here also the uncertainty principle
suffers some modification, but it has more to do with the discrete nature of
the spectrum than with this particular algebra. A similar situation already
arises for the uncertainty relation between angle and angular momentum with
eigenstates of angular momentum satisfying 
\begin{equation}
\Delta L_{z}\Delta \phi =0  \label{B1}
\end{equation}
in apparent contradiction with Eq.(\ref{10}). However it does not contradict
the (domain-correct) Eq.(\ref{10a}) which in this case is not equivalent to (%
\ref{10}). In fact, by integration by parts, what is obtained, instead of (%
\ref{10}), is\cite{Judge} \cite{Kraus} 
\begin{equation}
\Delta L_{z}\Delta \phi \geq \frac{1}{2}\left| 1-2\pi |\psi (2\pi
)^{2}|\right|  \label{B2}
\end{equation}

For the case $\ell \neq 0$ $\varepsilon =-1$, computing $\Delta x\Delta p$
for a localized state $e_{n}=\frac{1}{\sqrt{2\pi }}e^{-in\theta }$ one
obtains $\Delta x\Delta p=0$, which does not contradict Eqs.(\ref{10a}) or (%
\ref{10}), because $\left( e_{n},\cos \theta e_{n}\right) =0$.

This calculation however fail to convey the true physical meaning of the
uncertainty relations which should be a statement about the minimal size of
the phase-space cell that must be assigned to a quantum state. Therefore, a
formulation of the uncertainty principle that applies both to continuous and
to discrete spectrum may use, instead of the product $\Delta x\Delta p$, the
product of the inverses of the density of states $\mu \left( x\right)
^{-1}\mu \left( p\right) ^{-1}$. For example, for a free particle quantized
in a box of size $L$ , the density of momentum eigenstates is $\mu \left(
p\right) =\frac{L}{2\pi }$ and for each one of these states the density of
particles in a unit length is $\mu \left( x\right) =\frac{1}{L}$. Then 
\[
\mu \left( x\right) ^{-1}\mu \left( p\right) ^{-1}=2\pi 
\]
a reasonable statement about the average size of the phase-space cell.

On the other hand for continuous spectrum, $\mu \left( x\right)
^{-1}\thicksim \Delta x$ , $\mu \left( p\right) ^{-1}\thicksim \Delta p$ and
the uncertainty principle would have its usual meaning.

For the algebra (\ref{2}) with $\varepsilon =-1$ the density of eigenstates $%
e^{-in\theta }$ of the position operator is $\mu \left( x\right) =\ell ^{-1}$%
. On the other hand the density of states in momentum space is obtained by
integrating Eq.(\ref{7}) over a unit interval around $p=\left\langle
p\right\rangle =0$ leading to $\mu \left( p\right) =\frac{\ell }{\pi }$.
Then 
\[
\mu \left( x\right) ^{-1}\mu \left( p\right) ^{-1}=\pi 
\]

\end{document}